\documentclass[12pt]{article}
\usepackage{epsfig}

\newcommand{\be}{\begin{eqnarray}}
\newcommand{\ee}{\end{eqnarray}}

\def\mref#1{(\ref{#1})}

\def\bd{\begin{displaymath}}
\def\ed{\end{displaymath}}

\def\ba#1{\begin{array}{#1}}
\def\ea{\end{array}}
\def\nn{\nonumber}
\newfont{\Bbb}{msbm10 scaled 1200}

\begin{document}

\pagestyle{empty}

\begin{center}

{\LARGE\bf  Simple applications of fundamental solution method in 1D
   quantum mechanics\\A talk given at the conference on:\\ ``Geometry
   Symmetry and Mechanics I''\\Lisbon, 11-16 July 2001}
\vskip 18pt

{\large {\bf Stefan Giller}}

\vskip 3pt

Theoretical Physics Department II, University of {\L}\'od\'z,\\
Pomorska 149/153, 90-236 {\L}\'od\'z, Poland\\ 
e-mail: sgiller@krysia.uni.lodz.pl
\end{center}
\vspace{6pt} 
\begin{abstract}A method of fundamental solutions has been used to
  show its effectiveness in solving some well known problems of 1D
  quantum mechanics (barrier penetrations, over-barrier reflections,
  resonance states), i.e. those in which we look for exponentially small
  contributions to semiclassical expansions for considered quantities.
  Its usefulness for adiabatic transitions in two energy level systems
  is also mentioned.
\end{abstract}
\vskip 9pt

{\small PACS number(s): 03.65.-W , 03.65.Sq , 02.30.Lt , 02.30.Mv}

{\small Key Words: fundamental solutions, semiclassical expansion,
  JWKB approximations, exponential asymptotics}

\newpage

\pagestyle{plain}

\setcounter{page}{1}

\section*{I. Introduction}

\hskip+2em The semiclassical methods belong to
the most effective approximate methods in quantum theory \cite{1}. Starting as
the well known JWKB approximations \cite{22} the methods have been
succesively developed in terms of the wave function formalism by Fr\"oman and Fr\"oman in one dimension and
next extended to arbitrary finite dimension by Russian school of
Maslov, Fedoriuk and collaborators \cite{8,9,2}. Parallelly, a formulation
of the semiclassical methods relied on the Feynman path integral
formalism in quantum mechanics \cite{30} has been developed by
Gutzwiller \cite{29}. 

As it is well known semiclassical series expansions most frequently are
asymptotic expansions i.e. the corresponding series are
divergent \cite{20}. Therefore these series to be used have to be
abbreviated. This is usually done keeping only few first terms of the
series. However the best results are obtained if these abbrevations are
done at the {\it least} terms of the series. 

Nevertheless, in general, even in
the best cases there
are still some discrepencies left between the abbreviated series and the
aproximated quantities. It is well known that a source for these
discrepencies is related with additional, ``exponentially small''
contributions which are lost when the main semiclassical series are
constructed. Therefore, in order to improve accuracy of the
semiclassical methods it is necessary to recover these tiny
contributions \cite{11,31,6},\cite{12}-\cite{15}. 

Sometimes it is unavoidable. Namely there are cases
when the {\it main} contributions are built {\it uniquely} from these
exponentially small objects, i.e. the corresponding semiclassical
series vanish identically in these cases. The well known examples of
the latter situation are all
barrier penetration amplitudes \cite{1}. Therefore there is a need for
a systematic way of semiclassicall expansions containing also their
exponentially small elements.

One of ways used for this goal is the technique of Borel
resummation of semiclassical series \cite{5}. To use, however, the
latter one has to have solutions of the Schr\"odinger equation which
can be semiclassically expanded and which can be also recoverd from their
expansions by the method of Borel resummation. In other case the
problem of exponentially small contributions can not be well
defined. 

An existence of solutions with Borel summable semiclassical expansions
is not common - only limited sets of them can have this property. In fact,
an existence of such solutions is well established only in 1D quantum
mechanics \cite{10}. This is why a discussion which follows is limited to one
dimension. 

{\bf{\it A solution of a stationary Schr\"odinger equation which semiclassical expansion
is Borel summable to the solution itself is called a fundamental
solution.}}

For a given one dimensional potential there is a definite (finite or
infinite) set of fundamental solutions \cite{6,10,4}.

In the next section a construction of fundamental solutions is described.

\section*{II. Fundamental solutions and their properties}

\hskip+2em {\bf 1. Fundamental solutions }
\vspace{12pt}

Let us describe shortly basic lines in defining fundamental 
solutions \cite{8,9,10,4}. In the following we shall limit ourselves to one
dimensional potentials $V(x,\hbar)$ which have unique complex
extensions both into $x$- and $\hbar$-planes and which are meromorphic on
the $x$-plane while having the following asymptotic expansions in
$\Re\hbar>0$ for $\Re\hbar\to +\infty$:
\begin{eqnarray}
V(x,\hbar)\sim V_0(x)+\hbar V_1(x)+\hbar^2 V_2(x)+\ldots
\label{II.1}
\end{eqnarray}

Let us put $q(x,\hbar)=V(x,\hbar)-E$, where $E$ is an energy of the
quantum system and let $Z$ denote a set of all the points of the
$x$-plane at which $q(x,\hbar)$ has its single or double poles.
Let $\delta(x)$ be a meromorphic function of $x$, the unique singularities of which are double poles at
the points collected by $Z$ with coefficients at all the poles equal to $1/4$ each. (The latter
function can be constructed in general with the
help of the Mittag-Leffler theorem \cite{17}).

     Consider now a function
\begin{eqnarray}
\tilde{q}(x,\hbar)=q(x,\hbar)+\hbar^2\delta(x)
\label{3.1}
\end{eqnarray}

     A presence and a role of the $\delta$-term in (\ref{3.1}) are explained below. This term contributes
to (\ref{3.1}) if and only when a corresponding function $q(x,\hbar)$ contains simple or
second order poles. (Otherwise a corresponding $\delta$-term is put to zero). It is called a Langer
term \cite{10,4,18}.

     The Stokes graph corresponding to a function $\tilde{q}(x,\hbar)$ consists now of Stokes lines
emerging from roots (turning points) of $\tilde{q}(x,\hbar)$. Stokes lines satisfy one of the following
equations: 
\begin{eqnarray}
\Re\int_{x_i}^x\sqrt{\tilde{q}(\xi,\hbar)}d \xi=0
\label{3.2}
\end{eqnarray}
with $x_i$ being a root of $\tilde{q}(x,\hbar)$. We shall assume further a generic situation when all roots
$x_i$ are simple.

     Stokes lines which are not closed end at these points of the $x$-plane (i.e. have the latter points
as their boundaries) for which the action integral in (\ref{3.2}) becomes infinite. Of course such points
are singular for $\tilde{q}(x,\hbar)$ and they can be its finite poles or its poles lying at infinity. 

     Each such a singularity $z_k$ of $\tilde{q}(x,\hbar)$ defines a domain called a sector. This is the
connected domain of the $x$-plane bounded by Stokes lines and $z_k$
itself. The latter is also
a boundary for Stokes lines or an isolated boundary point of a sector (as it is in the
case of the second order pole). 

     In each sector the LHS in (\ref{3.2}) is only positive or only negative. 

     Consider now the Schr\"odinger equation:
\begin{eqnarray}
\Psi^{\prime\prime}(x) - \hbar^{-2} q(x) \Psi(x) = 0
\label{II.3}
\end{eqnarray}
corresponding to a
     potential $V(x,\hbar)$ and energy $E$ (we have put a mass $m$ in (\ref{II.3}) to be equal 
to $1/2$). 

Following Fr\"oman and
     Fr\"oman \cite{8} and Fedoriuk \cite{9} (see also \cite{10,4}) in each sector $S_k$ having a singular point $z_k$ at its
boundary one can
define a solution to (\ref{II.3}) having the following Dirac form:
\begin{eqnarray}
\Psi_{k}(x) = \tilde{q}^{-\frac{1}{4}}(x){\cdot}
e^{\frac{\sigma}{\hbar} W(x)}{\cdot}{\chi_{k}(x)} &
& k = 1,2,\ldots
\label{4}
\end{eqnarray}
where:
\begin{eqnarray}
\chi_{k}(x) = 1 + \sum_{n{\geq}1}
\left( -\frac{\sigma \hbar}{2} \right)^{n} \int_{z_{k}}^{x}d{\xi_{1}}
\int_{z_{k}}^{\xi_{1}}d{\xi_{2}} \ldots 
\int_{z_{k}}^{\xi_{n-1}}d{\xi_{n}}
\omega(\xi_{1})\omega(\xi_{2}) \ldots \omega(\xi_{n}) 
\label{5}
\end{eqnarray}
\begin{eqnarray*}
{\times} 
\left( 1 - e^{-\frac{2\sigma}{\hbar}{(W(x)-W(\xi_{1}))}} \right)
\left(1 - e^{-\frac{2\sigma}{\hbar}{(W(\xi_{1})-W(\xi_{2}))}} \right)
\cdots
\left(1 - e^{-\frac{2\sigma}{\hbar}{(W(\xi_{n-1})-W(\xi_{n}))}} \right)
\end{eqnarray*}
with
\begin{eqnarray}
\omega(x) = \frac{\delta(x)}{\tilde{q}^{\frac{1}{2}}(x,\hbar)} - 
{\frac{1}{4}}{\frac{\tilde{q}^{\prime\prime}(x,\hbar)}
{\tilde{q}^{\frac{3}{2}}(x,\hbar)}} +
{\frac{5}{16}}{\frac{\tilde{q}^{\prime 2}(x,\hbar)}
{\tilde{q}^{\frac{5}{2}}(x,\hbar)}}
\label{6}
\end{eqnarray}
and
\begin{eqnarray}
W(x) = \int_{x_{i}}^{x} \sqrt{\tilde{q}(\xi,\hbar)}d\xi
\label{7}
\end{eqnarray}
where $x_i$ is a root of $\tilde{q}(x,{\hbar^2})$ lying at the 
boundary of $S_k$.

In (\ref{4}) and (\ref{5}) a sign $\sigma\; (= \pm 1)$ and an
integration path are chosen in such a way to have: 
\begin{eqnarray}
\sigma \Re \left(W(\xi_{j}) - W(\xi_{j+1}) \right) \geq 0
\label{8}
\end{eqnarray}
for any ordered pair of integration variables (with $\xi_{0} = x$). 
Such a path of integration is then called {\it canonical}.  

The Langer $\delta$-term appearing in (\ref{6}) and (\ref{7}) 
is necessary to make all integrals in (\ref{5}) converging when $z_k$ is a first or a second order pole of 
 $\tilde{q}(x,{\hbar^2})$ 
or when solutions (\ref{4}) are to be continued to such poles.
As it follows from \mref{6} each such a pole $z_k$ demands a 
contribution to $\delta(x)$ 
of the form $(2(x-z_k))^{-2}$, what has been already assumed in
the corresponding construction of $\delta(x)$.

A domain $D_k (\supset S_k)$ where $\chi_k (x)$ can be
represented by (\ref{5}) with canonical integration paths is
called {\it canonical}.

\vspace{12pt}
\hskip+2em {\bf 2. Standard semiclassical expansion} \cite{19}
\vspace{12pt}

\hskip+2em Let us note that the $\chi$-factors entering the Dirac forms (\ref{4}) are 
the solutions of the following two second order linear differential
equations obtained by the substitution (\ref{4}) into the 
Schr\"odinger equation (\ref{II.3}): 
\begin{eqnarray}
 \tilde{q}^{- \frac{1}{4}}(x) \left( {\tilde{q}^{- \frac{1}{4}}(x)} \chi(x) 
\right)^{\prime{\prime}}
+\frac{2\sigma}{\hbar} \chi^{\prime}(x)+\tilde{q}^{-\frac{1}{2}}(x)\delta(x)\chi(x) = 0
\label{9}
\end{eqnarray}

Eqs.(\ref{9}) provide us with a general form of
semiclassical expansions for a $\chi$-factors if such expansions exists. 
Namely, assuming the latter we can substitute into (\ref{9}) 
the semiclassical expansion for $\chi$: 
\begin{eqnarray}
\chi(x) \sim\sum_{n\geq{0}} 
\left(-\frac{\sigma{\hbar}}{2} \right)^{n}
\kappa_{n}(x)
\label{10}
\end{eqnarray}
to get the following recurrent relations for $\kappa_n (x)$: 
\begin{eqnarray}
\kappa_n (x) = C_n +
\int_{x_{n}}^{x} \tilde{D}(y) \kappa_{n-1}(y) dy  
& \mbox{  ,  } \mbox{} n \geq 1 
\label{11}
\end{eqnarray}
where a {\it linear operator} $\tilde{D}(x)$ is given by:
\begin{eqnarray}
\tilde{D}(x) =  \tilde{q}^{-\frac{1}{4}}(x,\hbar)
\frac{d^2}{dx^2} \tilde{q}^{-\frac{1}{4}}(x,\hbar) +
\tilde{q}^{-\frac{1}{2}}(x,\hbar)\delta(x)
\label{12}
\end{eqnarray}
and
\begin{eqnarray*}
\kappa_0 (x) \equiv C_0 
\end{eqnarray*}
and where $x_n$, $n \geq 1$, are
arbitrary chosen regular points of $\omega (x)$ and $C_n$, $n \geq 0$, are
arbitrary constants. It is, however, easy to show that choosing
all points $x_n$ to be the same, say $x_0$, merely redefines constants $C_n$. Assuming this we get for $\kappa_n (x)$:
\begin{eqnarray}
\kappa_n (x) = \sum_{k=0}^{n} C_{n-k} I_k (x,x_0)
\label{13}
\end{eqnarray}
where
\begin{eqnarray*}
I_0 (x,x_0) \equiv  1
\end{eqnarray*}
and
\begin{eqnarray}
I_{n}(x,x_0) = \int_{x_0}^{x}d\xi_{n}\tilde{D}(\xi_{n}) 
\int_{x_{0}}^{\xi_{n}}d\xi_{n-1}\tilde{D}(\xi_{n-1}) 
\ldots \int_{x_0}^{\xi_3} d\xi_2 \tilde{D}(\xi_2)\int_{x_{0}}^{\xi_2}d\xi_{1}\omega(\xi_1)
\label{14}
\end{eqnarray}
\begin{eqnarray*}
n = 1,2,\ldots
\end{eqnarray*}

Substituting (\ref{13}) into (\ref{10}) we get finally for the
considered semiclassical expansion: 
\begin{eqnarray}
\chi (x,\hbar) \sim \chi^{as}(x,\hbar)\equiv\sum_{n\geq{0}} 
\left(-\frac{\sigma\hbar}{2} \right)^{n} C_n
\sum_{k\geq{0}} 
\left(-\frac{\sigma\hbar}{2} \right)^{k} I_k (x,x_0)
\label{15}
\end{eqnarray}

The form (\ref{15}) is called a {\it standard form} for the expansion (\ref{10}).

The formula (\ref{15}) can be applied
to fundamental solutions $\chi$- factors $\chi_k(x,\hbar)$ of (\ref{5}).
In this case however, due to the explicite form (\ref{5}) of these factors, 
their 
asymptotic expansions in corresponding canonical domains $D_k$ can be 
found directly from (\ref{5}) to be:
\begin{eqnarray*}
\chi_k (x,\hbar) \sim \chi_{k}^{as} (x,\hbar) =
1 + \sum_{n\geq{1}} 
\left(-\frac{\sigma_k {\hbar}}{2} \right)^{n} \kappa_{k,n}(x) 
\end{eqnarray*}
\begin{eqnarray}
\kappa_{k,n}(x)=I_n (x,z_k) 
\label{16}
\end{eqnarray}
where $z_k$ is a singular point in a corresponding sector $S_k$ where the 
solution (\ref{5}) is defined. 

The latter formula can be brought to the standard form (\ref{15}) by noticing 
that the following identity holds:
\begin{eqnarray}
\chi_{k,n}(x) =
\sum_{p=0}^n  \chi_{k,p}(x_0) I_{n-p} (x,x_0)
\label{17}
\end{eqnarray}
so that we get
\begin{eqnarray}
\chi^{as}(x,\hbar)=\chi^{as}(x_0,\hbar)
\sum_{k\geq{0}} 
\left(-\frac{\sigma\hbar}{2} \right)^{k} I_k (x,x_0)
\label{18}
\end{eqnarray}
i.e. a "constant" $C(\hbar)\equiv\sum_0^\infty
\left(-\frac{\sigma\hbar}{2} \right)^{n} C_n$ coincides with 
$\chi^{as}(x_0,\hbar)$ in this case.

\vspace{12pt}
\hskip+2em {\bf 3. Borel summability of semiclassical
    expansions} \cite{19}
\vspace{8pt}

\hskip+2em The standard semiclassical expansions (\ref{15}) can be tried to 
be Borel summed. 
Necessary and sufficient conditions for that have been 
formulated by Watson-Nevan-\\linna-Sokal theorem \cite{5}. In this respect, 
the following two important facts have been shown for the case of 1D quantum 
mechanics with meromorphic potentials \cite{19}:

{\bf $1^0$ Fundamental solution semiclassical expansions as defined 
by (\ref{16}) are Borel summable to these solutions themselves;

$2^0$ Fundamental solutions are the unique ones with the above property,
i.e. each semiclassical series (\ref{16}) which can be Borel summed 
(such a summation {\it always} provides us with a solution to the 
Schr\"odinger equation (\ref{II.3})) is {\it always} summed to some 
fundamental solution (up to a multiplicative constant).}

\section*{III. Examples of applications of fundamental solutions}

\hskip+2em Let us make a general statement that having for 
a given one dimensional problem a complete set of the corresponding 
fundamental solutions we can get an answer, exact as well as approximate, for 
any 
basic quantum mechanical question one can put for this problem. A typical list
of them is, as one knows, the following:

\hskip+2em{\bf 1. Quantization of bound states} \cite{10}
\vspace{12pt}

\hskip+2em{\bf 2. One dimensional scattering amplitudes: reflection and 
transmission coefficients}
\vspace{12pt}

\hskip+2em{\bf 3. Quantization of resonant states}
\vspace{12pt}

However, each time whenever higher dimensional problems can be reduced to one dimensional
ones the above questions can be solved effectively also for such cases. 
Typical here are problems with spherical or other symmetries among which 
the Coulomb or Yukawa potentials are well known examples. 

Since fundamental solutions are immanently related to JWKB and, wider, 
semiclassical expansions then they give rice to following further 
applications:

\hskip+2em {\bf 4. JWKB formulae for wave functions, energy 
levels, scattering amplitudes and matrix elements} \cite{10,7}
\vspace{12pt}

\hskip+2em {\bf 5. Exactness of JWKB formulae for energy level 
quantization} \cite{4}
\vspace{12pt}

\hskip+2em {\bf 6. Exponential asymptotics in semiclassical 
expansions} \cite{6}
\vspace{12pt}

We know also that a time evolution of two energy level systems 
can {\it always} be rephrased as the stationary Schr\"odinger equation 
with a time as the corresponding independent variable. Therefore the method 
of fundamental solutions can be applied also to:

\vspace{12pt}
\hskip+2em{\bf 7. Adiabatic transitions in two energy level systems} \cite{21}
\vspace{12pt} 

To illustrate some of these applications we shall consider a potential of 
the form $V(x)=\frac{x^2-1}{(x^2+1)^2}$ and the Coulomb potential. 
The latter - for comparison with known methods used in this case.

\vspace{18pt}
\hskip+2em{\bf a) Particle in a field of the potential
  $V(x)=\frac{x^2-1}{(x^2+1)^2}$}
\vspace{12pt}

\vskip 12pt
Fig.1 A potential $V(x)=\frac{x^2-1}{(x^2+1)^2}$
\vskip 12pt

  This potential is shown on Fig.1. 
It has two maxima at the points
  $x_{\pm}^{max}=\pm\sqrt{3}$ with a value $V(\pm\sqrt 3)=\frac{1}{8}$ and a
  minimum at $x^{min}=0$ with a value $V(0)=-1$. It has also two real roots
  at $x_{\pm}^{root}=\pm 1$ and two second order poles at $x_\pm
  ^{pole}=\pm i$. Therefore, depending on its energy $E$, a particle
  can be bounded by this potential for $-1<E<0$, can form
  resonant states for $0<E<\frac{1}{8}$ and can scatter for $0<E$,
  penetrating both the barriers for $0<E<\frac{1}{8}$ or being
  ``softly'' reflected in its over-barrier scattering for
  $E>\frac{1}{8}$.

For describing all the above possibilities in terms of fundamental
solution method we have to draw first Stokes graphs corresponding to
these cases. Since the potential considered has two second order poles
it has to acquire two
additional Langer terms, one for each pole, so that we have to draw Stokes graphs for the
following $\tilde q$-function:
\begin{equation}
\tilde
q(x,\hbar)=\frac{x^2-1}{(x^2+1)^2}-\frac{\hbar^2}{4}\left(\frac{1}{(x-i)^2}+\frac{1}{(x+i)^2}\right)-E
\label{19}
\end{equation}

Depending on an energy $E$ these graphs are shown on figures
2-4. They corresponds to different states of a particle and we
consider them consecutively.

\vspace{12pt}
{\it $1^0$ Bound states: $-1<E<0$}
\vspace{12pt}

\vskip 12pt
Fig.2 Stokes graph corresponding to energy range $-1<E<0$
\vskip 12pt

Fig.2 shows the ``first'' sheet (from an infinite number of them) of
a Riemann surface on which fundamental solutions are defined. A 
number of the latter is also infinite. $S_1,S_3,S_{\bar 3}, S_2$
denote the corresponding sectors lying on the ``first'' sheet, so that there are four
fundamental solutions defined on this sheet:
$\Psi_1(x),\Psi_3(x),\Psi_{\bar 3}(x)=\bar\Psi_3(\bar x),\Psi_2(x)$,
corresponding to the respective sectors. 

The solutions $\Psi_1(x)$
and $\Psi_2(x)$ vanishing at infinities of the respective sectors
$S_1,S_2$ (i.e. at
$\pm\infty$ respectively) can represent two 
branches of a corresponding bound state wave function. To do this they
have to coincide (up to a constant) with each other, i.e. we have:
\begin{equation}
\Psi_1(x)\equiv C\Psi_2(x)
\label{20}
\end{equation}

Writing (\ref{20}) first for $x=x_+^{pole}\in S_3$ and next for $x=
x_-^{pole}\in S_{\bar 3}$ we find:
\begin{eqnarray}
C=\frac{\chi_{1\to
      3}(\hbar,E)}{\chi_{2\to 3}(\hbar,E)}e^{-\frac{1}{2}\oint_K\sqrt{\tilde q(x,\hbar,E)}dx}=-\frac{\chi_{1\to\bar
      3}(\hbar,E)}{\chi_{2\to \bar3}(\hbar,E)}e^{+\frac{1}{2}\oint_K\sqrt{\tilde q(x,\hbar,E)}dx}
\label{50}
\end{eqnarray}
where an integration contour $K$ is shown on Fig.2 and $\chi_{1\to 3}$, etc. mean integrations in (\ref{5})
performed between
singular points of the corresponding sectors along canonical paths.

From (\ref{50}) we get immediately the following quantization condition
for energy levels:
\begin{eqnarray}
e^{-\frac{1}{\hbar}\oint_K\sqrt{\tilde q(x,\hbar,E)}dx}=-\frac{\chi_{1\to \bar 3}(\hbar,E)\chi_{2\to 3}(\hbar,E)}{\chi_{1\to
      3}(\hbar,E)\chi_{2\to {\bar 3}}(\hbar,E)}\nn\\
\\
=-\frac{\chi_{1\to \bar 3}(\hbar,E)\chi_{2\to 3}(\hbar,E)}{\overline{\chi_{1\to \bar 3}(\hbar,E)\chi_{2\to 3}(\hbar,E)}}\nn
\label{21}
\end{eqnarray}
with its obvious $JWKB$-limit: 
\begin{eqnarray}
e^{-\frac{1}{\hbar}\oint_K\sqrt{ q(x,\hbar,E^{JWKB})}dx}=-1
\label{51}
\end{eqnarray}

\vspace{12pt}
{\it $2^0$ Scattering amplitudes - reflection and transmission
  coefficients: $0<E<\frac{1}{8}$}
\vspace{12pt}

\vskip 12pt
Fig.3 Stokes graph corresponding to energy range $0<E<\frac{1}{8}$
\vskip 12pt

A corresponding graph is shown on Fig.3. We assume the plane
wave is incoming from the left. It reflects mainly from the barriers
but also penetraits them both but with much smaller
amplitude. Therefore for $x$ sufficiently close to $+\infty$ there is
only outgoing plane wave while for $x$ sufficiently close to $-\infty$
there are two waves: incoming and reflected. 

The outgoing wave can be
represented by the fundamental solution $\Psi_1(x)$ since at $x$ close
to $+\infty$ it becomes a plane wave with positive momentum. 

Similarly, the solution $\Psi_2(x)$ can represent the incoming wave
while the solution $\Psi_{\bar 2}(x)$ - the reflected one.

It is clear that to solve our problem we have to find a relation:
\begin{equation}
\Psi_2(x)= R\Psi_{\bar2}(x)+T\Psi_1(x)
\label{22}
\end{equation}
where the reflection amplitude $R$ and the transmission one $T$ have to be expressed only in terms
of fundamental solution factors. 

Note, that (\ref{22}) expresses not only a
physical fact but also a mathematical one that every solution of the 
Schr\"odinger equation is a linear combination of another two if the
latter are linear independent and the linear independence of any two
fundamental solutions is their basic property.

As previously,  Eq.(\ref{22}) has  to be valid  for {\it any}  $x$ and
therefore  it means that  the rhs  of (\ref{22})  is an  {\it analytic
continuation}  of  $\Psi_2(x)$.   However,  we have  to  perform  this
analytic continuation  being constrained  that it goes  entirely along
canonical paths. Such a continuation ensures keeping the full controll
over  the  semiclassical  properties   of  the  amplitudes  $R$  and
$T$. This continuation  can be done in many ways  but the simplest one
is to  take the solution  $\Psi_1(x)$ (instead of $\Psi_2(x)$)  and to
continue it canonically to the sector $S_2$ or $S_{\bar 2}$.

There  are two  groups of  sectors  in Fig.2 such that in each group
its sectors can contact
canonically with each other. These are $S_1,S_{\bar 1},S_3,S_{\bar 3}$
and $S_2,S_{\bar 2},S_3,S_{\bar  3}$. Therefore continuing $\Psi_1(x)$
we express  it first by a  linear combination of  $S_3,S_{\bar 3}$ and
next the latter two solutions - as linear combinations of $S_2,S_{\bar
  2}$. We get:

\begin{eqnarray}
\Psi_1(x)=\alpha_{\frac{1}{3}\to\bar
  3}\Psi_3(x)+\alpha_{\frac{1}{\bar 3}\to 3}\Psi_{\bar3}(x)\nn\\
\nn\\
\Psi_3(x)=\alpha_{\frac{3}{2}\to\bar
  2}\Psi_2(x)+\alpha_{\frac{3}{\bar 2}\to 2}\Psi_{\bar2}(x)\\
\nn\\
\Psi_{\bar3}(x)=\overline{\Psi_3(\bar x)}\nn
\label{23}
\end{eqnarray}
where
\begin{eqnarray} 
\alpha_{\frac{i}{j}\to
  k}=\lim_{x\to z_k}\frac{\Psi_i(x)}{\Psi_j(x)}
\label{24}
\end{eqnarray}
and $x$ goes to a singular point $z_k$ of the sector $S_k$ along a
canonical path.

Finally we get
\begin{eqnarray}
\Psi_1(x)=\left(\alpha_{\frac{1}{3}\to\bar
  3}\alpha_{\frac{3}{2}\to\bar
  2}+\alpha_{\frac{1}{\bar 3}\to 3}\alpha_{\frac{\bar3}{ 2}\to \bar2}\right)\Psi_2(x)+\left(\alpha_{\frac{1}{3}\to\bar
  3}\alpha_{\frac{3}{\bar2}\to 2}+\alpha_{\frac{1}{\bar 3}\to
  3}\alpha_{\frac{\bar 3}{\bar 2}\to\ 2}\right)\Psi_{\bar2}(x)
\label{25}
\end{eqnarray} 

Compairing (\ref{25}) with (\ref{22}) we obtain:

\begin{eqnarray}
R=-\frac{\alpha_{\frac{1}{3}\to\bar
  3}\alpha_{\frac{3}{\bar2}\to 2}+\alpha_{\frac{1}{\bar 3}\to
  3}\alpha_{\frac{\bar 3}{\bar 2}\to\ 2}}{\alpha_{\frac{1}{3}\to\bar
  3}\alpha_{\frac{3}{2}\to\bar
  2}+\alpha_{\frac{1}{\bar 3}\to 3}\alpha_{\frac{\bar3}{ 2}\to
  \bar2}}\nn\\
\\
T=\frac{1}{\alpha_{\frac{1}{3}\to\bar
  3}\alpha_{\frac{3}{2}\to\bar
  2}+\alpha_{\frac{1}{\bar 3}\to 3}\alpha_{\frac{\bar3}{ 2}\to
  \bar2}}\nn
\label{26}
\end{eqnarray}

All $\alpha$-coefficients in the above formulae can be easily expressed in
terms of the three factors constituting the Dirac form (\ref{4}) of
fundamental solutions to get:

\begin{eqnarray}
R=i\frac{\chi_{1\to\bar3}\chi_{2\to 3}+\chi_{1\to 3}\chi_{2\to \bar3}e^{\frac{2}{\hbar}\int_{-x_1}^{x_1}\sqrt{\tilde
      q(x)}dx}}{\chi_{1\to3}\chi_{\bar2\to\bar
      3}+\chi_{1\to\bar 3}\chi_{\bar2\to 3}e^{\frac{2}{\hbar}\int_{-x_1}^{x_1}\sqrt{\tilde
      q(x)}dx}}\nn\\
\\
T=-\frac{\chi_{3\to\bar 3}}{\chi_{1\to 3}\chi_{\bar2\to\bar
      3}e^{-\frac{1}{\hbar}\int_{-x_1}^{x_1} \sqrt{\tilde
      q(x)}dx}+\chi_{1\to\bar 3}\chi_{\bar2\to 3}e^{\frac{1}{\hbar}\int_{-x_1}^{x_1}\sqrt{\tilde
      q(x)}dx}}e^{\frac{1}{\hbar}\oint_{K_2}\sqrt{\tilde
      q(x)}dx} \nn
\label{27}
\end{eqnarray}

A semiclassical limit which follows from (\ref{27}) is immediate:
\begin{eqnarray}
R^{JWKB}=i\nn\\
\\
T^{JWKB}=-\frac{1}{2\cos\left(\frac{1}{\hbar}\int_{-x_1}^{x_1} \sqrt{
      q(x)}dx\right)}e^{\frac{1}{\hbar}\oint_{K_2}\sqrt{
      q(x)}dx}\nn
\label{28}
\end{eqnarray}

The JWKB formula for $T$ seems to suggest a singular behaviour of 
this amplitude for energies for which cosine in its denominator is 
close to zero. This is, in fact, a situation closely related to the
problem discussed in the next point and, as one knows, discribes so
called resonant scattering. Of course, in the last case the above
$JWKB$-formula for $T$ has to be modified leading us to its Breit -
Wigner form.

\vspace{12pt}
{\it $3^0$ Resonant states - resonance widths}: $0<E<\frac{1}{8}$
\vspace{12pt}

For the energy range as given above the
potential considered gives rice to 
resonant states to exist. A condition for their
existence follows directly from Eq.(\ref{25}) as a demand for the
solution $\Psi_2(x)$ not to contribute to this equation, i.e. resonant wave functions can have only branches {\it outgoing} to both
infinities of the real $x$-axis. Therefore a corresponding condition
takes the form:
\begin{eqnarray}
\alpha_{\frac{1}{3}\to\bar
  3}\alpha_{\frac{3}{2}\to\bar
  2}+\alpha_{\frac{1}{\bar 3}\to 3}\alpha_{\frac{\bar3}{ 2}\to \bar2}=0
\label{29}
\end{eqnarray}
i.e. resonant energies have to be singular points (poles) for both the
amplitudes $R$ and $T$. 

In terms of $\chi$-factors this condition reads:
\begin{eqnarray}
e^{-\frac{2}{\hbar}\int_{-x_1}^{x_1} \sqrt{\tilde
      q(x)}dx}=-\frac{\chi_{1\to\bar 3}\chi_{\bar2\to 3}}{\chi_{1\to 3}\chi_{\bar2\to\bar
      3}}
\label{30}
\end{eqnarray}

Eq.(\ref{30}) is very similar to (\ref{21}). The main difference between these 
formulae lies, however, in energies for which they can be satisfied. For the 
latter these energies are real (and negative), for the former the corresponding 
energies can be only complex.

To see this let us note that $\alpha$-coefficients are, in general, not 
independent. Despite a trivial relations like the following 
$\alpha_{\frac{i}{j}\to k}\alpha_{\frac{j}{i}\to k}=1$ there are non 
trivial ones which follow from the fact that among any four fundamental 
solutions which can communicate canonically with themselves only a pair of 
them are linear independent. This limitation leads to the following 
{\it identity} \cite{10}:
\begin{eqnarray}
\alpha_{\frac{i}{j}\to k}=\alpha_{\frac{i}{j}\to l}+\alpha_{\frac{i}{l}\to j}
\alpha_{\frac{l}{j}\to k}
\label{31}
\end{eqnarray}
for each quartet $i,j,k,l$.

For the case considered there are, as we have already mentioned, two groups of 
fundamental solutions, each containing four of them, related to the sectors 
$S_1,S_{\bar 1},S_3,S_{\bar 3}$ and $S_2,S_{\bar 2},S_3,S_{\bar  3}$. Therefore 
there are the following two identities for corresponding $\alpha$-coefficients:
\begin{eqnarray}
\alpha_{\frac{1}{\bar 1}\to 3}=\alpha_{\frac{1}{\bar1}\to \bar3}+
\alpha_{\frac{1}{\bar 3}\to \bar1}
\alpha_{\frac{\bar3}{\bar1}\to 3}\nn\\
\\
\alpha_{\frac{2}{\bar 2}\to 3}=\alpha_{\frac{2}{\bar2}\to \bar3}+
\alpha_{\frac{2}{\bar 3}\to \bar2}
\alpha_{\frac{\bar3}{\bar2}\to 3}\nn
\label{32}
\end{eqnarray}
which can be written by $\chi$-factors as:
\begin{eqnarray}
\chi_{1\to 3}\chi_{\bar1\to\bar 3}=\chi_{\bar1\to 3}\chi_{1\to\bar 3}-\chi_{3\to\bar 3}
e^{-\frac{1}{\hbar}\oint_{K_1}\sqrt{\tilde q(x)}dx}\nn\\
\\
\chi_{2\to 3}\chi_{\bar2\to\bar 3}=\chi_{\bar2\to 3}\chi_{2\to\bar 3}-\chi_{3\to\bar 3}
e^{\frac{1}{\hbar}\oint_{K_2}\sqrt{\tilde q(x)}dx}\nn
\label{33}
\end{eqnarray}

The last equations are written, by assumption,  for a {\it complex} energy $E$. 
However, for a real one they take the forms:
\begin{eqnarray}
|\chi_{1\to 3}|^2=|\chi_{\bar1\to 3}|^2-\chi_{3\to\bar 3}e^{-\frac{1}{\hbar}\oint_{K_1}\sqrt{\tilde q(x)}dx}\nn\\
\\
|\chi_{2\to 3}|^2=|\chi_{\bar2\to 3}|^2-\chi_{3\to\bar 3}e^{\frac{1}{\hbar}\oint_{K_2}\sqrt{\tilde q(x)}dx}\nn
\label{34}
\end{eqnarray}

Now, we use the last identities to show that putting $E=E_0-i\frac{\Gamma}{2}$ we get a 
width $\Gamma$ as exponentially small quantity in comparison with $E_0$. 
To see this let us write the "quantization" condition (\ref{30}) in the 
following form:
\begin{eqnarray}
e^{-\frac{2}{\hbar}\int_{-x_1}^{x_1} \sqrt{\tilde
      q(x)}dx}=-e^{i\phi_{1\to\bar 3}+i\phi_{\bar2\to 3}-i\phi_{1\to 3}-i\phi_{\bar2\to\bar
      3}}\left\vert\frac{\chi_{1\to\bar 3}\chi_{\bar2\to 3}}{\chi_{1\to 3}\chi_{\bar2\to\bar
      3}}\right|
\label{35}
\end{eqnarray}
where $\phi_{1\to\bar 3},\phi_{\bar2\to 3},\phi_{1\to 3},\phi_{\bar2\to\bar
      3}$ are phases of the corresponding $\chi$-factors.

We now put
\begin{eqnarray}
-\frac{2}{i\hbar}\int_{-x_1}^{x_1} \sqrt{\tilde
      q(x,E_0)}dx=(2k+1)\pi+\phi_{1\to\bar 3}+\phi_{\bar2\to 3}-\phi_{1\to 3}-\phi_{\bar2\to\bar
      3}
\label{36}
\end{eqnarray}
so that we get
\begin{eqnarray}
-\frac{2}{\hbar}\int_{-x_1}^{x_1} \sqrt{\tilde
      q(x,E_0-i\frac{\Gamma}{2})}dx+\frac{2}{\hbar}\int_{-x_1}^{x_1} \sqrt{\tilde
      q(x,E_0)}dx=\frac{1}{2}\ln\left\vert\frac{\chi_{1\to\bar 3}\chi_{\bar2\to 3}}{\chi_{1\to 3}\chi_{\bar2\to\bar
      3}}\right\vert_{E=E_0-i\frac{\Gamma}{2}}^2
\label{37}
\end{eqnarray}

We now expand both the sides of (\ref{37}) into
the Taylor series around $E_0$ abbraviating the latters on the first terms. We get
\begin{eqnarray}
\frac{i}{\hbar}\Gamma\frac{d}{dE_0}\int_{-x_1}^{x_1} \sqrt{\tilde
      q(x,E_0)}dx=\frac{1}{2}\ln\left(\left\vert\frac{\chi_{1\to\bar 3}\chi_{\bar2\to 3}}{\chi_{1\to 3}\chi_{\bar2\to\bar
      3}}\right\vert_{E=E_0}^2-i\frac{\Gamma}{2}\frac{d}{dE_0}\left\vert\frac{\chi_{1\to\bar 3}\chi_{\bar2\to 3}}{\chi_{1\to 3}\chi_{\bar2\to\bar
      3}}\right\vert_{E=E_0}^2\right)
\label{38}
\end{eqnarray}

Next, using (\ref{34}), we obtain
\begin{eqnarray}
\frac{2}{\hbar}\Gamma\int_{-x_1}^{x_1} \frac{1}{\sqrt{\left|\tilde
      q(x,E_0)\right|}}dx=\nn\\
\\
\ln\left[\left(1-i\frac{\Gamma}{2}\frac{d}{dE}\right)\left(1+\frac{\chi_{3\to\bar 3}}{|\chi_{1\to\
      3}|^2}e^{-\frac{1}{\hbar}\oint_{K_1}\sqrt{\tilde
      q(x)}dx}\right)\left(1+\frac{\chi_{3\to\bar 3}}{|\chi_{2\to\
      3}|^2}e^{\frac{1}{\hbar}\oint_{K_2}\sqrt{\tilde
      q(x)}dx}\right)\right]_{E=E_0}\nn
\label{39}
\end{eqnarray}

It follows from (\ref{39}) that, up to higher order terms in
exponentials, $\Gamma$ is given by
\begin{eqnarray}
\Gamma=\frac{\hbar}{2\int_{-x_1}^{x_1} \frac{1}{\sqrt{\left|\tilde
      q(x,E_0)\right|}}dx}\left(\frac{\chi_{3\to\bar 3}(E_0)}{|\chi_{1\to\
      3}(E_0)|^2}+\frac{\chi_{3\to\bar 3}(E_0)}{|\chi_{2\to\
      3}(E_0)|^2}\right)e^{\frac{1}{\hbar}\oint_{K_2}\sqrt{\tilde
      q(x,E_0)}dx}
\label{40}
\end{eqnarray}

Finally, in a JWKB limit we get
\begin{eqnarray}
\Gamma^{JWKB}=\frac{2\hbar}{T}e^{\frac{1}{\hbar}\oint_{K_2}\sqrt{
      q(x,E_0)}dx}
\label{41}
\end{eqnarray}
where $T$ is a classical period of a particle moving between the two barriers.

\vspace{12pt}
{\it $4^0$ Over barrier scattering amplitudes: $\frac{1}{8}<E$}
\vspace{12pt}

A Stokes graph corresponding to this case is shown on Fig.4. A
scattering wave function is represented as previously by the fundamental solution
$\Psi_2$, a reflected one - by $\Psi_{\bar2}$ and a transmitted one -
by $\Psi_1$. As in the previous case these solutions are related by
Eq.(\ref{22}) and the reflection and transmission amplitudes are given
again by Eqs.(\ref{26}). A difference which has to appear between
these two cases lies of course in formulae expressing
$\alpha$-coefficients by corresponding $\chi$-factors. For the case
considered we get according to Fig.4:
\begin{eqnarray}
R=i\frac{\chi_{1\to\bar3}\chi_{2\to3}+\chi_{1\to3}\chi_{2\to\bar3}}{\chi_{1\to\bar3}\chi_{\bar2\to3}+\chi_{1\to3}\chi_{\bar2\to\bar3}e^{\frac{2}{\hbar}\int_{{\bar
      x}_1}^{x_1}\sqrt{\tilde
      q(x)}dx}}e^{\frac{1}{\hbar}\int_{{\bar x}_1}^{x_1}\sqrt{\tilde
      q(x)}dx}\nn\\
\\
T=\frac{\chi_{3\to\bar3}e^{-\frac{1}{\hbar}\int_{x_1}^{-{\bar x}_1}\sqrt{\tilde
      q(x)}dx}}{\chi_{1\to\bar3}\chi_{\bar2\to3}+\chi_{1\to3}\chi_{\bar2\to\bar3}e^{\frac{2}{\hbar}\int_{{\bar
      x}_1}^{x_1}\sqrt{\tilde
      q(x)}dx}}\nn
\label{42}
\end{eqnarray}

\vskip 12pt
Fig.4 Stokes graph corresponding to energy range $\frac{1}{8}<E$
\vspace{20pt}

where an integration between the points $x_1,-{\bar x}_1$ can run
along the Stokes line linking these points. Therefore the corresponding integral is
purely imaginary. 

On the other hand, an integral between the points ${\bar
      x}_1,x_1$ is {\it real} and {\it negative} since it can
    be taken along the {\it anti-Stokes} line linking these points. 

JWKB approximations to the above amplitudes follows immediately to be:
\begin{eqnarray}
R^{JWKB}=ie^{\frac{1}{\hbar}\int_{{\bar x}_1}^{x_1}\sqrt{
      q(x)}dx}\nn\\
\\
T^{JWKB}=e^{-\frac{1}{\hbar}\int_{x_1}^{-{\bar x}_1}\sqrt{
      q(x)}dx}\nn
\label{43}
\end{eqnarray}

\vspace{12pt}
\hskip+2em{\bf b) Particle in a Coulomb field}
\vspace{12pt}

Let $R_l(r),\; l=0,1,...$ , denote radial parts of the Coulomb wave
functions corresponding to subsequent orbital momentum
numbers. As it is well known functions $\Psi^{(l)}(r)=rR_l(r),\;
l=0,1,...$ , satisfy then
Schr\"odinger equations (\ref{II.3}) with respective potentials
$V_l(r)=-\frac{\alpha}{r}+\frac{\hbar^2l(l+1)}{r^2},\;l=0,1,...$, shown
in Fig.5, and with
boundary conditions $\Psi^{(l)}(0)=0,\; l=0,1,...$, .

\vskip 12pt
Fig.5 Coulomb potential corrected by centrifugal term
\vskip 12pt

To apply the fundamental solution formalism to two problems considered
in this case, i.e. to bound states ($E<0$) and to a scattering problem
($E>0$) the additive Langer term $\frac{\hbar^2}{4r^2}$ has to be
introduced to each potential $V_l(r)$ to give $\tilde
{V}_l(r)=-\frac{\alpha}{r}+\frac{\hbar^2\left(l+\frac{1}{2}\right)^2}{r^2},\;l=0,1,...$,
. Then figures 6 and 7 show respective Stokes graphs corresponding to
the above two problems. 

\vskip 12pt
Fig.6 Coulomb Stokes graph corresponding to energy range $E<0$ 
\vskip 12pt

\vspace{12pt}
\hskip+2em{\it $1^0$ Bound state energies}: $E<0$
\vspace{12pt}

Fig.6 shows the Stokes graph corresponding to this case. The
fundamental solutions
$\Psi_1^{(l)}(r)$ and $\Psi_3^{(l)}(r)$ corresponding to the sectors
$S_1$ and $S_3$ respectively are these which satisfy necessary
boundary conditions: the first one vanish at $r=+\infty$, the second -
at $r=0$. Therefore, identifying them with each other we obtain a wave function of a
bound state and a condition for energies of this state. We have:
\begin{eqnarray}
\Psi_1^{(l)}(r)=C_l\Psi_3^{(l)}(r)\nn\\
\\
C_l=\frac{e^{-\frac{1}{2\hbar}\oint\sqrt{\tilde V_l -E}dx}}{\chi_{2\to3}^{(l)}}=-\frac{e^{+\frac{1}{2\hbar}\oint\sqrt{\tilde V_l -E}dx}}{\chi_{\bar2\to3}^{(l)}}\nn
\label{44}
\end{eqnarray}

\vskip 12pt
Fig.7 Coulomb Stokes graph corresponding to energy range $E>0$
\vspace{12pt}

The last equation provides us simultanuously with the corresponding
quantization condition. If we perform necessary integrations in
the $\chi$-factors contained in this equation along the {\it negative}
real $r$-axis then we note that both these factors are {\it real}. On
the other hand they are mutually {\it complex conjugated}. Therefore, they
have {\it to cancel} each other, so that we get the following net
result for the quantization condition in the Coulomb potential:
\begin{eqnarray}
\frac{i}{\hbar}\oint\sqrt{\tilde V_l -E}dx=(2k+1)\pi,\;\;\;\;\; k=0,1,...
\label{45}
\end{eqnarray}

The above formula is a famous JWKB formula of Langer which appeared to be
{\it exact}. As we can see this is not accidental.

\vspace{12pt}
\hskip+2em{\it $2^0$ Coulomb scattering}: $E>0$
\vspace{12pt}

This situation is represented by Fig.7 where the solution $\Psi_{\bar1}^{(l)}(r)$ 
describes an {\it incoming} wave while $\Psi_1^{(l)}(r)$ - an {\it outgoing} 
one. It is now sufficient only to write such a combination of both these wave 
functions to make it vanishing at $r=0$. But this property is the one of the
solution $\Psi_3^{(l)}(r)$, i.e. the combination mantioned has to be 
proportional to the last solution. But it means that to find a corresponding 
scattering amplitude $S_l$ it is enough to express $\Psi_{\bar1}^{(l)}(r)$ as 
a linear combination of $\Psi_1^{(l)}(r)$ and $\Psi_3^{(l)}(r)$. We get from 
Fig.7:
\begin{eqnarray}
\Psi_{\bar1}^{(l)}(r)=\frac{\chi_{\bar1\to3}^{(l)}}{\chi_{1\to3}^{(l)}}\Psi_1^{(l)}(r)+
\frac{1}{\chi_{3\to1}^{(l)}}\Psi_3^{(l)}(r)
\label{46}
\end{eqnarray}

It follows from (\ref{46}) that
\begin{eqnarray}
S_l=\frac{\chi_{\bar1\to3}^{(l)}}{\chi_{1\to3}^{(l)}}=e^{2i\phi_{\bar1\to3}^{(l)}}
\label{47}
\end{eqnarray}
where $\phi_{\bar1\to3}^{(l)}$ is a phase of $\chi_{\bar1\to3}^{(l)}$. Its 
JWKB approximation follows directly from (\ref{47}) to be:
\begin{eqnarray}
\phi_{\bar1\to3}^{(l),JWKB}= \frac{\hbar}{2}\Im\int_0^\infty\omega(r)rdr\nn\\
\nn\\
\omega(r)=\frac{1}{4r^2\tilde{q}^{\frac{1}{2}}(r,\hbar)} - 
{\frac{1}{4}}{\frac{\tilde{q}^{\prime\prime}(r,\hbar)}
{\tilde{q}^{\frac{3}{2}}(r,\hbar)}} +
{\frac{5}{16}}{\frac{\tilde{q}^{\prime 2}(r,\hbar)}
{\tilde{q}^{\frac{5}{2}}(r,\hbar)}}\\
\nn\\
\tilde q(r,\hbar)=-\frac{\alpha}{r}+\frac{\hbar^2(l+\frac{1}{2})^2}{r^2},\;\;\;\;\;\;l=0,1,...
\nn
\label{48}
\end{eqnarray}
where an integration path is anyone which does not coincide with any of
the real half axes.

\section*{IV. Conclusions}

\hskip+2em As we have demonstrated the method of fundamental solutions
appears to be very effective in solving {\it any} well defined problem
of one-dimensional Schr\"odinger equation. It provides us both with {\it
exact} solutions of such problems and {\it immediately} with their
semiclassical approximations, including also exponentially small
corrections to main semiclassical series.

\end{document}